\documentclass[iop,apj]{emulateapj}

\usepackage{amsmath}

\newcommand{\kakkoi}[3]{\left(\frac{{#1}}{{#2}}\right)^{#3}}
\newcommand{\BF}[1]{\mbox{\normalsize \boldmath $#1$}}
\newcommand{\e}[1]{\times 10^{#1}}
\newcommand{\bracket}[1]{\langle {#1} \rangle}

\shorttitle{Solid Accretion onto Circumplanetary Disks}
\shortauthors{Tanigawa et al.}

\begin{document}


\title{Accretion of Solid Materials onto Circumplanetary Disks from Protoplanetary Disks}


\author{Takayuki Tanigawa}
\affil{Institute of Low Temperature Science, Hokkaido University,
    Sapporo, 060-0819, Japan}
\email{tanigawa@pop.lowtem.hokudai.ac.jp}

\and

\author{Akito Maruta and Masahiro N. Machida}
\affil{Department of Earth and Planetary Sciences, Kyushu University,
Fukuoka 812-8581, Japan}


%


\begin{abstract}
We investigate accretion of solid materials onto circumplanetary disks
from heliocentric orbits rotating in protoplanetary disks, which is a
key process for the formation of regular satellite systems.
%
In the late stage of gas-capturing phase of giant planet formation, the
accreting gas from protoplanetary disks forms circumplanetary disks.
Since the accretion flow toward the circumplanetary disks affects the
particle motion through gas drag force, we use hydrodynamic simulation
data for the gas drag term to calculate the motion of solid materials.
%
We consider wide range of size for the solid particles
($10^{-2}$-$10^6$m), and find that the accretion efficiency of the solid
particles peaks around 10m-sized particles because energy dissipation of
drag with circum-planetary disk gas in this size regime is most
effective.  The efficiency for particles larger than 10m size becomes
lower because gas drag becomes less effective.  For particles smaller
than 10m, the efficiency is lower because the particles are strongly
coupled with the back-ground gas flow, which prevent particles from
accretion.
We also find that the distance from the planet where the particles are
captured by the circumplanetary disks is in a narrow range and well
described as a function of the particle size.


\end{abstract}






\section{Introduction}
%
The giant planets in our solar system have many natural satellites.
In terms of mass, most of the satellites are categorized into regular
satellites, which are rotating in almost circular and co-planer with the
equatorial planes of the parent planets.
Because of the regularity, the satellites are believed to be formed in
circumplanetary disks, which would have existed when the giant planet
were forming in the protoplanetary disk.

Thus the satellite systems had been considered to have formed in an
isolated and closed disk that have enough mass to produce the current
systems \citep{Lunine82}.
This is so-called Minimum Mass Sub Nebula (MMSN) disk model.
However, the formation through such a heavy disk leads to some
difficulties in its formation processes, such as too high temperature
for H$_2$O to be solid phase, too fast type I migration for satellites,
and too short accretion timescale for Callisto's internal structure not
to be fully differentiated \citep{Canup02}.
In order to overcome these problems, two further models to describe
circumplanetary disks are proposed.
One is a gas-starved disk model \citep{Canup02, Canup06, Ward10}, which
is an open disk model.
This means that the disk receives continuous mass supply from the
protoplanetary disk, and is much less massive than the MMSN-type disk.
This model solves several serious problems that could not solve by
MMSN-type disk model \citep{Canup02}.
Another is solids-enhanced minimum mass (SEMM) model
\citep{Mosqueira03a, Estrada09}, which consists of a compact heavy
component and a wide-spread less massive one.
The two components are produced by the difference of specific angular
momentum of inflow gas, and the difference corresponds to whether gap
along the planet orbit exists or not.
The difference of the observed moment of inertia of Ganymede and
Callisto was tried to be explained by the large difference of surface
density between the two components.

The structure of a circumplanetary disk have been studied by
hydrodynamic simulations.
There are pioneering works that tried to see circumplanetary disks
\citep{Miki82, Sekiya87, Korycansky96},
and as computational speed became faster, the structure of the
circumplanetary disk became clearer by two-dimensional simulations
\citep{Kley99, Lubow99, Tanigawa02} with nested-grid method
\citep{DAngelo02} and three-dimensional simulations \citep{DAngelo03,
Bate03, Klahr06}.
In particular, recent simulations have revealed the circumplanetary disk
structure and the accretion flow onto the disk in very high resolution
with some special techniques, such as nested-grid methods in Eulerian
codes \citep{Machida08, Machida10, Paardekooper08, TOM12, Gressel13,
Szulagyi14} or SPH methods \citep{Ayliffe09b}, in addition to the recent
development of high-performance computers.

However, satellites around the giant planets are made of solid, and
supply of solid material into circumplanetary disks have not been
studied so far.
There are some studies that considered accretion of particles onto giant
planets under the influence of gas flow in protoplanetary disks for
dust- or boulder-size particles
\citep{Rice06,Paardekooper07,Ayliffe12,Zhu12} or planetesimals
\citep{Zhou07, Shiraishi08}, but the structure of gas flow near the
planet, such as circumplanetary disks, was not resolved in such studies.
In the phase of giant planet growth, circumplanetary disks are rotating
around the planet almost in Keplerian velocity, and the density would be
much higher than that in protoplanetary disks \citep{Ayliffe09b, TOM12}.
The particle motion is thus expected to be affected significantly by the
circumplanetary disks when they are captured, and high-resolution
structure of the gas flow near the planet is therefore necessary to be
considered.

In this study, we examine the supply of solid material onto the
circumplanetary disk by simulating motion of particles that are
originally rotating in heliocentric orbits.
In \S 2, we will explain formulation of our model, in \S 3 results of
orbital simulation will be shown, and we discuss issues that we do not
address in this paper and that might be important, and summarize our
results in \S 5.

\section{Methods}
%
We consider a growing giant planet embedded in a protoplanetary disk.
In the disk, particles in heliocentric orbits are rotating in the
protoplanetary disk.
In this study, we simulate the particle motion whether the particle are
captured by the circumplanetary disk under the influence of gas
accretion flow onto the giant planet.
We consider that the planet is rotating in a circular orbit with no
inclination from the midplane of the protoplanetary disk.

\subsection{Basic equations}
In order to investigate the orbits of particles around the planets in
detail, we use Hill's equation \citep[e.g.,][]{Henon86, Nakazawa88} with
a gas drag term.  Hill's equation describes motion of small particles
near a planet that is rotating around the central star, and adopts a
frame rotating with a planet that is static at the origin of the
coordinate on the frame.
Hill's equation is usually normalized by Hill's radius for length,
inverse of orbital angular velocity of the planet for time.  The
non-dimensional equation of the particles on the Hill coordinate can be
written as
\begin{equation}
\frac{d\tilde{\BF{v}}}{d\tilde{t}}
 = - \nabla \tilde{\Phi}
   - 2 \BF{e}_z \times \tilde{\BF{v}}
   + \tilde{\BF{a}}_{\rm drag},
\label{EOM}
\end{equation}
where $\BF{e}_z$ is unit vector in $z$-direction, $\tilde{\BF{v}}$ is
velocity, $\tilde{t}$ is time.  The second term in the right-hand side
is colioris force, which arises from the frame is rotating with the
planet orbital motion.  Normalized Hill potential $\tilde{\Phi}$ is
given by
\begin{equation}
\tilde{\Phi}
 = - \frac{3}{\tilde{r}}
   - \frac{3}{2} \tilde{x}^2
   + \frac{1}{2} \tilde{z}^2
   + \frac{9}{2},
\end{equation}
where $\tilde{r} = \sqrt{\tilde{x}^2 + \tilde{y}^2 + \tilde{z}^2}$.  The
first term in the right-hand side corresponds to the planet potential,
the second and third term describe tidal potential in holizontal and
vertical direction, respectively.  The last constant term is added so
that potential at the Lagrange points 1 and 2 becomes zero.  The
acceleration due to gas drag $\tilde{\BF{a}}_{\rm drag}$ is described by
\begin{equation}
\tilde{\BF{a}}_{\rm drag}
 \equiv
    \frac{\BF{F}_{\rm drag}/m}{r_{\rm H}\Omega_{\rm K}^2}
 = - \frac{3}{8} C_{\rm D}
     \frac{\rho_{\rm g}}{\rho_{\rm s}}
     \tilde{r}_{\rm s}^{-1} \Delta \tilde{u} \Delta \tilde{\BF{u}},
\label{a_drag_define}
\end{equation}
where $\BF{F}_{\rm drag} = (C_{\rm D}/2)\pi r_{\rm s}^2 \rho_{\rm g}
\Delta u \Delta \BF{u}$ is the drag force for a particle with radius
$r_{\rm s}$, $m$ is the mass of the particle, $C_{\rm D}$ is
non-dimensional gas drag coefficient, $\rho_{\rm g}$ is gas density,
$\rho_{\rm s}$ and $\tilde{r}_{\rm s}$ are the internal density and the
normalized physical radius of the particles, and $\Delta \BF{u}$ is the
velocity of the objects relative to the gas.  Variables with tildes
denote non-dimensional quantities.

\subsection{Effect of gas flow}
\subsubsection{Hydrodynamic simulation}
We use the gas flow that was obtained by \citet{TOM12}.  In order to
obtain gas flow with high resolution near the planet, they employed a
three-dimensional hydrodynamic simulation with a nested grid code
\citep{MachidaMTH05}, which was originally developed to explore the star
formation process by a collapse of the molecular cloud core
\citep{Matsumoto03b}.  The nested grid technique enables them to obtain
very high resolution gas flow in the vicinity of the planet.  In the
calculation they used 11 levels for nested grid.  They adopted Hill's
coordinate, which also contributes to enhance resolution near the
planet.

In their simulation, the ratio of Hill's radius to scale height of the
protoplanetary disk, which is the only one parameter of the system, was
adopted to be unity.  This corresponds to $M_{\rm p} \sim 120 M_{\rm
E}(a/5.2{\rm AU})^{3/4}$ for $T = 280 (a/{\rm 1AU})^{-1/2}$K, where
$M_{\rm E}$ is Earth mass and $a$ is semi-major axis of the planet.
The planet is assumed to be in the active gas accretion phase, which
corresponds to the stage after the onset of nucleated instability
\citep{Mizuno80, Bodenheimer86, Ikoma00}, but not to be embedded in a
very deep gap.

\subsubsection{Background gas flow}
\begin{figure*}
\epsscale{0.9}
\plotone{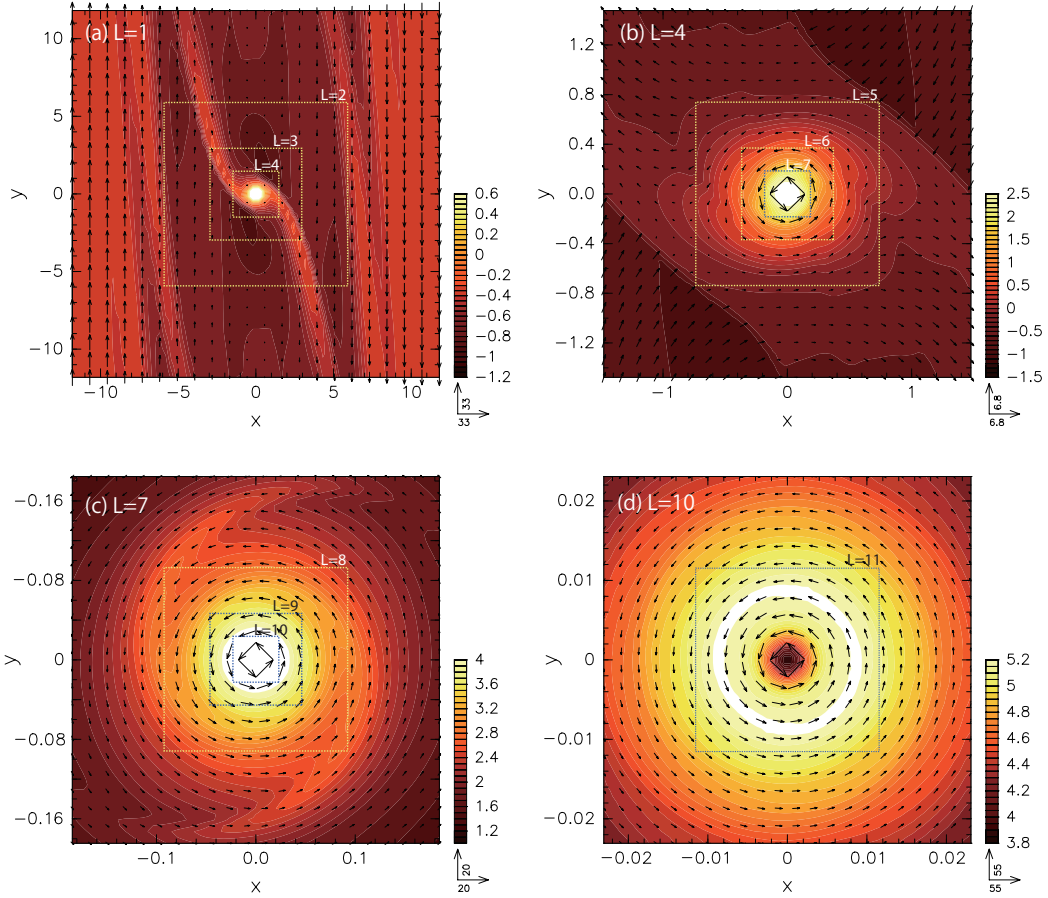}
%
\caption{Velocity (arrows) and log density (color) of the flow around
the planet on the midplane with four different nested level: $l=1$, 4,
7, 10, where difference of three in level means 8 ($=2^3$) times
difference in spatial scale.  Length of arrows are normalized by the two
arrows in the right bottom of the each panel.  Low density region near
the origin ($\tilde{r} \lesssim 0.005$) mainly arises from sink
treatment around the origin.
\label{fig:gas_flow}}
\end{figure*}

Fig.~\ref{fig:gas_flow} shows gas density and velocity field of the flow
at the midplane.  In Fig.~\ref{fig:gas_flow}a that showing wide field
flow mainly focusing on outside the Hill sphere, we can see two-arm
shock structure from the Hill sphere of the planet.  The shock structure
corresponds to spiral structure propagating in global (protoplanetary)
disks.
Fig.~\ref{fig:gas_flow}b shows the same flow but enlarged view focusing
around the Hill sphere.  We can see that there are shocks along the
lines through $(\tilde{x}, \tilde{y}) \sim (\pm 1.5, 0)$ and $(0,\pm
1.5)$, where gas has discontinuity in velocity and density.  Gas inside
the Hill sphere shows prograde rotation.
Fig.~\ref{fig:gas_flow}c shows even enlarged view.  In this scale ($\sim
0.1$ scale height), we can see another two-arm spiral structure around
the planet, but the non-axisymmetric structure disappears in even
smaller scale ($\sim 0.01$ scale height) as in Fig.~\ref{fig:gas_flow}d.
Note that low density region ($\tilde{r} \lesssim 0.005$) arises from
sink condition around the origin (see \citet{TOM12} in detail).

\subsubsection{Gas drag coefficient}
\label{sec:C_D}
\begin{figure}
\epsscale{1.0}
\plotone{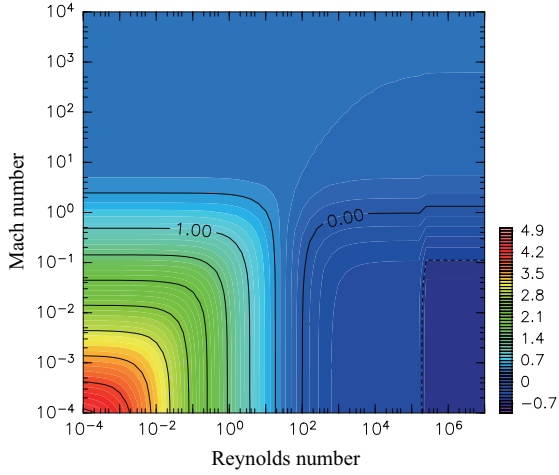}
\caption{Log of gas drag coefficient $C_{\rm D}$ (Eq.~\ref{eq:C_D}) as a
 function of the Reynolds number and the Mach number.
\label{fig:watanabe_C_D}}
\end{figure}

The gas drag coefficient $C_{\rm D}$ we adopt is an approximated formula
written in the form \citep{Watanabe97}:
\begin{equation}
C_{\rm D}
 \simeq
 \left[ \left( \frac{24}{{\cal R}} + \frac{40}{10+{\cal R}}
        \right)^{-1}
       +0.23 {\cal M}
 \right]^{-1}
 + \frac{(2.0-w){\cal M}}
        {1.6+{\cal M}}
 + w,
\label{eq:C_D}
\end{equation}
where the Reynolds number ${\cal R} = 2 r_{\rm s}u/\nu$, the Mach number
${\cal M}=u/c$, and $w$ is a correction factor depending on the Reynolds
number; $w=0.4 ({\cal R}<2\e{5})$ and $w=0.2 ({\cal R}>2\e{5})$.
Relative velocity between gas and particles is $u$, $c$ is isothermal
sound speed, $\nu$ is kinetic viscosity $\nu = 0.353 \sqrt{8/\pi} c
\ell_{\rm g}$ \citep{CC70}, $\ell_{\rm g}$ is mean free path
\footnote{We define mean free path as $\ell_{\rm g} = m_{\rm
mol}/(\sigma_{\rm mol} \rho_{\rm g})$, where $m_{\rm mol}$ and
$\sigma_{\rm mol}$ are mass and collision cross section of molecule,
whereas \citet{CC70} defined it as $\ell_{\rm g} = m_{\rm mol}/(\sqrt{2}
\sigma_{\rm mol} \rho_{\rm g})$, which makes apparent difference of the
coefficients in the formulae of viscosity.}.
Fig.~\ref{fig:watanabe_C_D} shows the value of $C_{\rm D}$ as a function
of the Mach number and the Reynolds number.

As in Eq.~(\ref{eq:C_D}), $C_{\rm D}$ is a function of the two
non-dimensional numbers: the Mach number and the Reynolds number.
However we need the ratio of the particle size to mean free path of
molecules when we evaluate the Reynolds number.  Thus we convert the
result of hydrodynamic simulation, which is obtained in non-dimensional
form, into quantities with real dimensions.
To do that, we adopt a disk model for gas temperature $T = 280 {\rm K}
(a/{\rm 1AU})^{-1/2}$ and gas surface density $\Sigma_{\rm g} = 1.7\e{4}
f_{\rm H}\, {\rm kg/m}^2 (a/{\rm 1AU})^{-3/2}$, where $f_{\rm H}$ is
scaling factor relative to that of the minimum mass disk model
\citep{Hayashi85}.  We adopt $\sigma_{\rm mol} = 2.0\e{-19}$ m$^2$ and
$m_{\rm mol} = 3.9\e{-27}$kg.  We fixed $a = 5.2$ AU in this paper and
fiducial value for $f_{\rm H}$ is 1.

\subsection{Numerical method}
We integrate Eq.~(\ref{EOM}) for particles with wide range of size,
using the Runge-Kutta-Fehlberg method with adaptive step size
\citep[e.g.,][]{numerical_recipes}.
We consider a two-dimensional problem; the orbits of particles is in the
same plane of the planet orbit and the midplane of the protoplanetary
disk.  We also restrict ourselves to initially zero-eccentricity
particles.
Because of these simplification, we only have one parameter; impact
parameter $\tilde{b}$, which is defined as the value of $x$ coordinate
of the particle position at $\tilde{y} \rightarrow \infty$.  In
numerical simulation, we cannot set infinite $\tilde{y}$ as an initial
position of the particles, thus we set the initial position
$(\tilde{x}_0, \tilde{y}_0)$ where $\tilde{x}_0^2 = \tilde{b}^2 -
8/\tilde{y}_0$, which is valid when $\tilde{x}_0 \ll \tilde{y}_0$
\citep{Ida89, Ohtsuki99}.  We set $\tilde{y}_0 = 100$ and $\tilde{x}_0$
is less than 3, so the double inequalities are met in our case.

The termination conditions of the orbital integration are follows:
(1) Collision with the planet.  We terminate numerical integration when
$\tilde{r} < \tilde{r}_{\rm p}$ where $\tilde{r}_{\rm p}$ is physical
size of the planet in our unit.  We set $\tilde{r}_{\rm p}= 0.001$,
which roughly corresponds to the physical size of a planet at 5AU.  In
the gas-free case case, results depends on the size of the planet, but
we mainly focus on the case where particles are captured by the
circumplanetary disks, not by the planet, so the physical size of the
planet is not important in this work as long as the size is small
enough.
(2) Receding from the planet: $|\tilde{y}| > \tilde{y}_0$.  These
particles first approach at least about a few Hill's radii and then move
away from the planet without collision with the planet or captured by
the circumplanetary disk.
\begin{figure*}
%
%
\epsscale{0.9}
\plotone{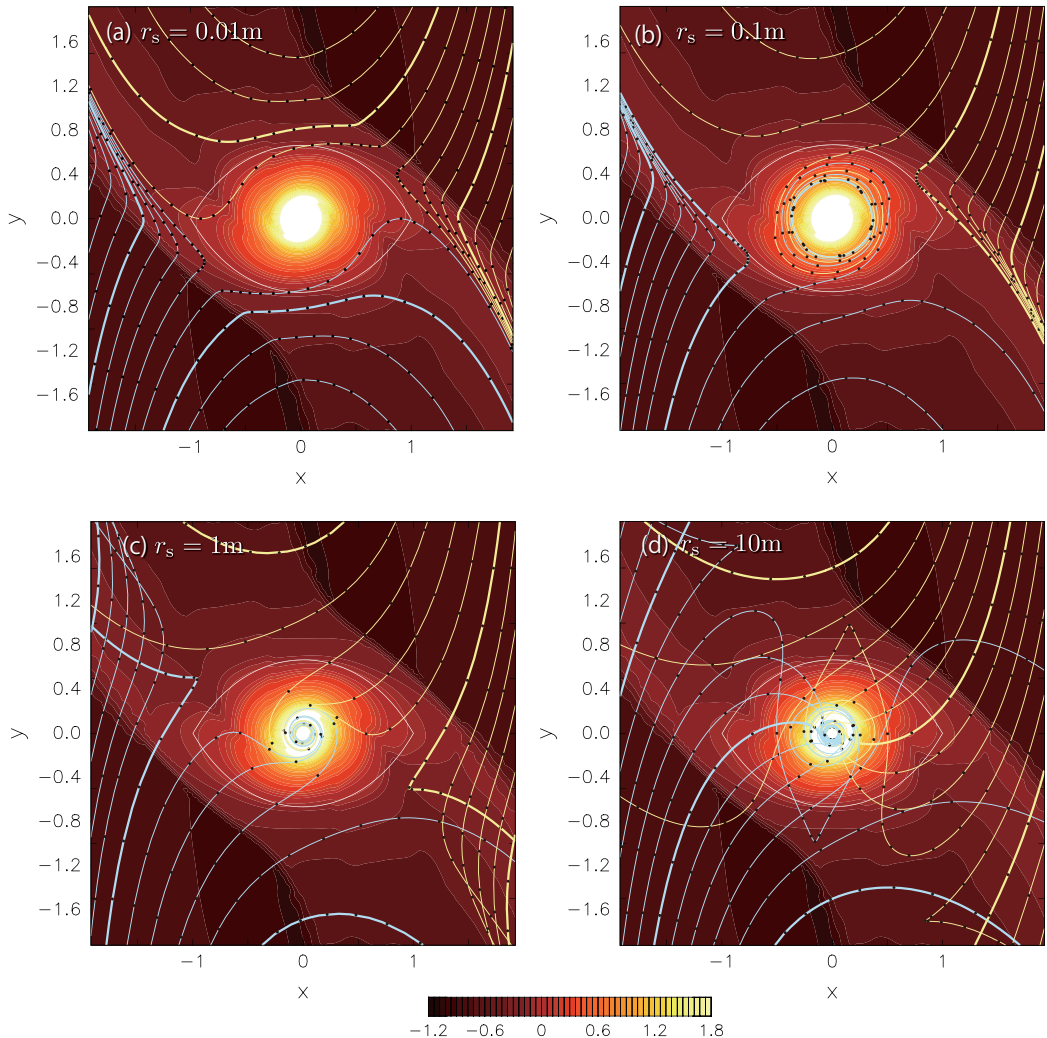}
%
\caption{Orbits of particles with size $r_{\rm s} = 10^{-2}$m
$10^{-1}$m, $10^0$m, $10$m.  Background colors show log10 of gas
density, and yellow and blue lines show orbits of the particles.  Black
dots on the orbits are put every 0.2 unit time.  White line shows the
Hill sphere.
\label{fig:orbit_small}}
\end{figure*}

As described above, the particles are assumed to be on the midplane
of the protoplanetary disk and initially in a circular orbit around the
central star.
We consider wide range of particle size, so this assumption would not be
always valid, but particles in a size range in which accretion to the
circumplanetary disk is effective (see Section 3) can be considered to
be settled down toward the midplane even when we consider stirring up of
particles by turbulence.
The thickness of the solid particles $h_{\rm d}$ is given by
\citep{Okuzumi12, Youdin07}
\begin{equation}
h_{\rm d}
 = h
   \left( 1 + \frac{\Omega t_{\rm s}}{\alpha}
              \frac{1 + 2\Omega t_{\rm s}}
                   {1 +  \Omega t_{\rm s}}
   \right)^{-1/2}
\end{equation}
where $\Omega$ is angular velocity of Keplerian rotation around the
central star, $\alpha$ is non-dimensional turbulent viscous parameter
\citep{Shakura73}, $t_{\rm s}$ is stopping time of particles.
%
If we assume $\alpha \sim 10^{-2}$, thickness of 1m-sized particles
layer is 1/10 of scale height of the gas disk at 5AU, and the typical
size for effective accretion is roughly larger than 1 meter, as we will
see, so the two-dimensional approximation is reasonable.
Once particles are in a thin layer, inclination cannot be pumped up by
gravitational scattering.
On the other hand, eccentricity is easier to be enhanced by the planet
gravity \citep{Ida90,Ohtsuki02}.  If synodic period is longer than the
stopping time, the assumption of circular orbit should be valid because
eccentricity would be damped until the next approach by Keplerian shear.
This condition roughly corresponds to the size $r_{\rm s} \lesssim 100$
m.  Particles with sizes larger than $\sim 100$m, however, would have
some eccentricity comparable to the order of unity when they approach
the planet, which would affect the result.  Although we should keep this
in mind, we assume circular orbit for the initial condition of the
particles for simplicity.

\section{Results}
\subsection{Typcal orbits of captured particles}
\subsubsection{Strong gas drag case: Orbits of small particles}
We first describe particle motion in the case of strong gas drag, which
corresponds to particles with size roughly smaller than 1m.
%
Fig.~\ref{fig:orbit_small} shows orbits of small particles
($r_{\rm s} \leq 10$m) around the Hill sphere.
%
Fig.~\ref{fig:orbit_small}a shows orbits of 1cm-sized, which is almost
the same as streamline of gas because gas and particles are well coupled.
Gas in the region $\tilde{x}>0$ approaches with Keplerian shear motion in
negative $y$ direction from large $\tilde{y}$ region.  For gas that
closes with the Hill sphere passes the shock surface that enhances
density and reduces velocity.  Gas that reaches at about
$(\tilde{x},\tilde{y}) \sim (1.0, 0.5)$ bifurcates toward two streams in
front of the Hill sphere; one crosses the $y$-axis (the planet orbit),
makes U-turn, and goes back to positive $y$ direction.  The other stream
passes by the Hill sphere without crossing $y$-axis and moves towards
negative $y$ direction (see Fig.~\ref{fig:gas_flow} and also
\citet{TOM12} in detail).
Since gas in the midplane does not accrete onto the circumplanetary
disk, 1cm particles do not either.
For 10cm-sized particles (Fig.~\ref{fig:orbit_small}b), overall orbits
outside the Hill sphere looks very similar to that of 1cm-sized case,
but one clear difference is that there are orbits that enters the Hill
sphere and accretes into the circumplanetary disk, although gas does not
enter it through the midplane.  This is because, although the particles
are well coupled with gas in the Keplerian timescale, the particles just
after the shock surface tend to decouple with gas in a short timescale,
which leads to the deviation of the orbit from gas motion.  Near the
bifurcation point of the gas flow and in front of the shock surface, the
motion of the particles is directed toward the planet, which enable the
particles to intrude into the Hill sphere against the drag of gas that
is not going to enter.
This feature becomes more significant for larger particles.  In the case
of 1m-sized particles (Fig.~\ref{fig:orbit_small}c), there is wider
band in which the particles are accreting onto the circumplanetary disk.
This means that the deviation of the particle motion from the gas flow
is more significant especially after the shock surface.  In addition, we
can see orbits that cross with each other, which does not occur in
the case of smaller particles.  This is one of typical behaviors of
motion for decoupled particles.
In the case of particles with 10m (Fig.~\ref{fig:orbit_small}d), the
motion outside the Hill sphere is almost free from gas drag, but if the
particles go into the Hill sphere and get closer to the planet, the
particles are captured by the denser gas of the circumplanetary disk at
the deeper region.
Note that there is a orbit that looks like deflected at $(x,y) \simeq
(0.2, 1.0)$ is an apparent motion on the rotating frame.  The particle
are actually rotating smoothly on the inertial frame even around the
apparent deflected point, but the Hill coordinate are rotating with the
Keplerian angular velocity of the planet orbital motion, and the
rotating velocity is subtracted on the Hill coordinate.  Thus the orbit
looks like deflected.  This feature is notable where the distance from
the planet is near the Hill radius because, in that region, the
Keplerian angular velocity around the planet is close to that around the
central star.

\subsubsection{Weak gas drag case: Orbits of large particles}
\begin{figure}
\epsscale{1.0}
\plotone{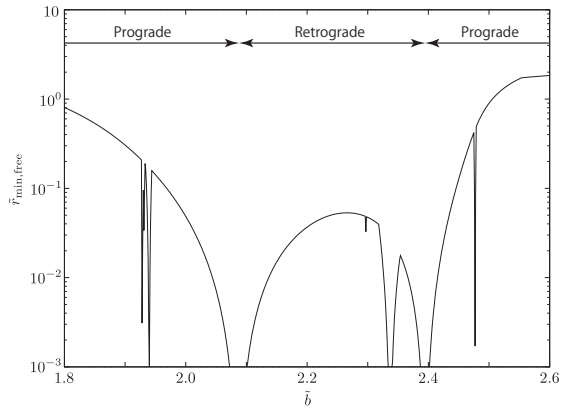}
\caption{Minimum distance to the planet of particles with initially
 circular and no inclination orbits as a function of impact parameter
 $\tilde{b}$ in the gas-free case ($\tilde{r}_{\rm min,free}$).  In the
 regions $\tilde{b}\lesssim 2.1$ and $\tilde{b} \gtrsim$ 2.4, particles
 encounter the planet in prograde direction, while $2.1 \lesssim
 \tilde{b} \lesssim 2.4$ particles encounter it in retrograde direction.
\label{fig:rmin_vs_b}}
\end{figure}

Next we describe particle motion in the case of weaker gas drag, which
corresponds to particles with size roughly larger than 1m, although the
size ranges for the two cases (strong and weak gas drag cases) overlap
with each other, which promotes a deeper understanding of the capturing
process.
Before showing orbits of the particles, we introduce minimum distance to
the planet of an orbit as a function of impact parameter $\tilde{b}$ in
the gas free case.
%
Fig. \ref{fig:rmin_vs_b} shows minimum distance between particles from
the planet through the orbits for gas-free case, which was presented by
\citet{Petit86,Ida89}.  The distance is referred to as $\tilde{r}_{\rm
min,free}$ in this paper.
There are two main collisional bands \citep{Giuli68}, which divide
encounter type into three in terms of encounter direction; $\tilde{b}
\lesssim 2.1$, $2.1 \lesssim \tilde{b} \lesssim 2.4$, and $\tilde{b}
\gtrsim 2.4$, which correspond respectively to prograde, retrograde, and
prograde encounters.
We can expect that, in the retrograde encounter regime, particles tend
to get strong gas drag and are easy to be captured, while particles in
the prograde encounter regimes are more difficult to be captured.
Note that there are very narrow bands which show close encounter in a
discontinuity manner with respect to $\tilde{b}$, which arises from
multiple encounter in each orbit \citep[e.g.,][]{Nishida83, Ida89}.  But
this is so narrow that the bands does not have any significant effect on
the solid accretion rate onto the circumplanetary disk in a statistical
sense.

\begin{figure}
\epsscale{1.0}
\plotone{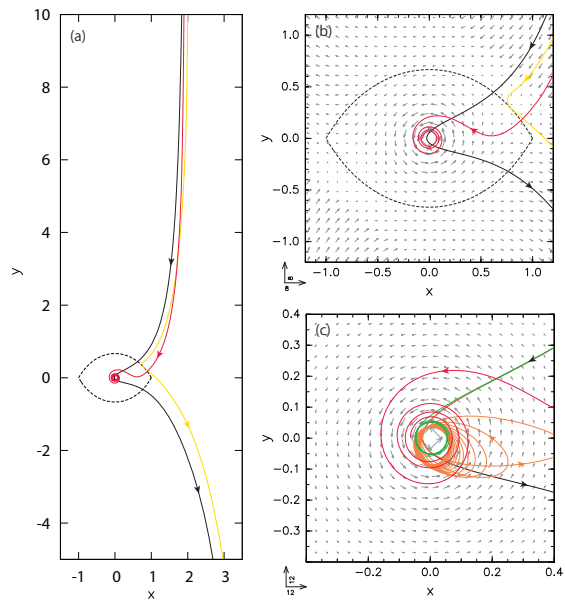}
%
%
\caption{Orbits of particles of $\tilde{b}=2.022$, which corresponds to
 prograde encounter region.  Left and right upper panels shows orbits of
 particles in the case with $r_{\rm s} = 0.1$m (yellow), 1m (red), and
 gas-free (black), respectively.  Right lower panel shows $r_{\rm s} =
 1$m (red), 100m (green), 10000m (Orange) cases.  Vectors in the middle
 and right panels shows velocity field of the gas flow.
\label{fig:orbit_b2022}}
\end{figure}

Fig.~\ref{fig:orbit_b2022} shows example orbits in prograde capturing
regime.  We show orbits of several particle sizes in the case with
$\tilde{b} = 2.022$, in addition to the gas-free case which corresponds
to the orbit of $\tilde{r}_{\rm min} = 0.0250$ (see also
Fig.~\ref{fig:rmin_vs_b}).
Fig.~\ref{fig:orbit_b2022}a shows orbits in wide area focusing on how
particles approach the Hill sphere from heliocentric orbits.  We cannot
see any significant difference between the three cases until they reach
the Hill sphere including the gas-free case.
Fig.~\ref{fig:orbit_b2022}b shows close-up view of
Fig.~\ref{fig:orbit_b2022}a.  We can see that the particle of 10cm size
does not enter the Hill sphere and recedes from it.  This is because the
particle is well coupled with gas, as mentioned before.
For the 1m-sized particle, it can penetrate into the Hill sphere through
the low velocity gas at the post shock region.  Although the deviation
of the orbit from the gas flow is sensible for the intruding motion, the
particle still get significant effect from the gas that rotates prograde
direction, thus the particle starts rotating also in the same direction.
Fig.~\ref{fig:orbit_b2022}c shows more close-up view of orbits of other
three different sizes (1m, 100m, 10000m).  In the case of 1m size, we
can see that the particle gradually spirals into inner region.
In the case of 100m size, the particle motion is almost the same as that
of gas-free case until distance from the planet becomes less than about
0.2.  But after the first encounter at $(\tilde{x}, \tilde{y}) = (-0.05,
0)$, the particle is immediately circularized in a few orbit around the
planet.  Once the orbit is circularized, the orbit does not change
because the gas motion in this region is almost circular, which results
in weak gas drag force.
In the case of 10000m size, the particle moves along with the orbit of
gas-free case even around the first encounter, but the particle is
captured in the Hill sphere because of energy dissipation by the gas
drag through the first encounter, and the orbit becomes highly
eccentric.  Since the gas drag is not so effective in comparison with
smaller particle cases, it takes longer time to be circularized.  In the
course of the circularization, the distance of apocenter continuously
decreases whereas the pericenter does not change significantly.

\begin{figure}
\epsscale{1.0}
\plotone{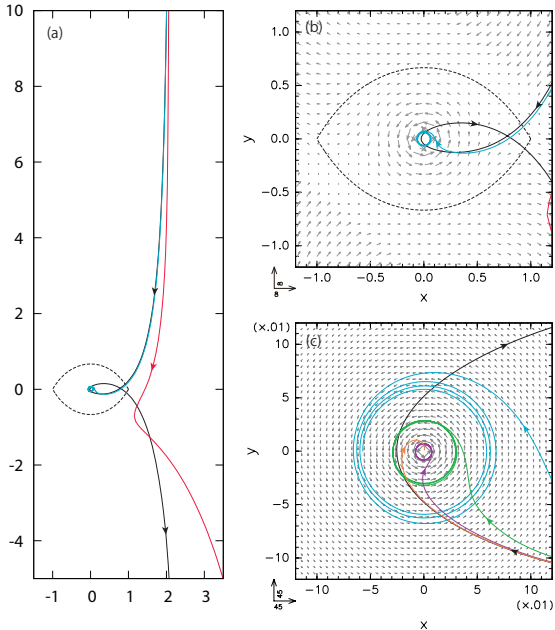}
%
%
\caption{Orbits of particles of $\tilde{b}=2.174$, which corresponds to
 retrograde encounter region.  Left and right upper panels show $r_{\rm
 s} = 1$m (red), 10m (blue), and gas free (black) cases.  Right lower
 panel shows $r_{\rm s} = 10$m (blue), 100m (green), 1000m (purple),
 10000m (orange) and gas-free cases.  Vectors in the right
 panels show velocity field of the gas flow.
\label{fig:orbit_b2174}}
\end{figure}

Fig. \ref{fig:orbit_b2174} shows orbits of particles in the retrograde
encounter regime ($\tilde{b}=2.174$) with several-size particles as well
as the gas-free case.  The distance at the closest approach for the
gas-free case is $\tilde{r}_{\rm min} = 0.0253$, which is similar to
that of the prograde case of $\tilde{b}=2.022$ shown in the above.
Fig. \ref{fig:orbit_b2174}a shows orbits in the wide field.  For the
gas-free case, the particle enters the Hill sphere and encounters with
the planet, then escapes from the Hill sphere.  For the small particles
($r_{\rm s}=$ 1m), the motion of approaching the Hill sphere is similar
to that of the gas-free case, but they cannot enter the Hill sphere
because of the strong gas drag with the gas that does not enter the Hill
sphere, which can also be observed in the prograde case.
Figs. \ref{fig:orbit_b2174}b and c show close-up views of orbits.
Unlike the 1m case, 10m and larger particles can enter the Hill sphere
across the high-density low-velocity region after the shock surface.
However, the motion of 10m-sized particle is strongly affected at
$\tilde{r} \lesssim 0.2$ by the motion of the gas that is in prograde
rotation.  Thus the 10-sized particle, which was originally moving in
retrograde direction, flips the direction to prograde, and rotates in
almost circular orbit in accordance with the motion of the
circumplanetary disk.  The spiral-in motion is due to the drag from the
gas, which is rotating in sub-Keplerian velocity.
In the case of larger particles (100m and 1000m), the tendency is
similar.  But the effect of the gas drag becomes weaker, so the point of
turn-over to prograde becomes closer to the planet.  For both cases, the
particles settle in circular orbits, and the orbital radius of
circularization decreases with increasing particle size.  The deviation
from true circles (inward spiral movement) is less significant compare
to that of the 10m case.
However, in the case of the 10000m-sized particle, the gas drag is so
weak that the particle cannot change the direction from retrograde to
prograde in the course of the approach to the planet, and falls to the
planet before circularization or change the direction in accord with gas
flow.

\subsection{Capture radius in circumplanetary disks}
\label{sec:captured_radius}
In order to consider processes of satellite formation in a
circumplanetary disk, we need to know where solid particles are supplied
at the circumplanetary disk.  As we showed in the previous section, the
captured particles eventually become circular orbits in the prograde
direction in a short timescale unless the particles collide with the
planet before being circularized.  Since the relative velocity with the
gas after the circularization is very small, the timescale of orbital
evolution due to gas drag becomes much longer than that of the
circularization.  We therefore define the captured radius as a distance
from the planet at the circularization, which is different from the
normal definition of capture in an energetic sense; Jacobi energy
$\tilde{E}_{\rm J} = \tilde{v}^2/2 + \tilde{\Phi}$ becomes negative.
More specifically, we define the captured radius at the time when either
of the two condition is met:
(1) Circularized in the circumplanetary disk: $\tilde{E}_{\rm J} < 0$
and $e<0.3$ and $\tilde{a}<0.5$ and $N_{\rm w} \geq 3$, where $e$ and
$\tilde{a}$ are eccentricity and semi-major axis of of the particle
around the planet, and $N_{\rm w}$ is winding number
\citep{Kary96,Iwasaki07}.  When a particle crosses the $x$- or $y$-axis
in the prograde direction around the planet, 1/4 is added to $N_{\rm
w}$, while the same amount is subtracted from $N_{\rm w}$ when it
crosses the axes in the retrograde direction.
(2) Winded capture: $\tilde{E}_{\rm J} < 0$, and $N_{\rm w} \geq 15$.
When one of the above conditions is met, the captured radius
$\tilde{r}_{\rm cap}$ is determined as the larger one of the two:
pericenter of the orbit at the time when the condition is met, or the
minimum distance from the planet until the time when the condition is
met.
We define $\tilde{r}_{\rm cap}$ as the larger one of $\tilde{a}$ or
$\tilde{r}_{\rm min}$ when either of the two conditions are met, and we
do not define it when neither of the two are met.
The former condition (1) is mainly for the weak gas-drag cases where the
orbit is gradually shifting toward circular from highly eccentric orbit.
The latter (2) is for the strong gas-drag cases where osculating
Keplerian orbital elements are difficult to determine.
Note that there are adjustable parameters to determine the captured
radius, but the result is not sensitive to the parameters.

\begin{figure*}
\epsscale{1.0}
%
\plotone{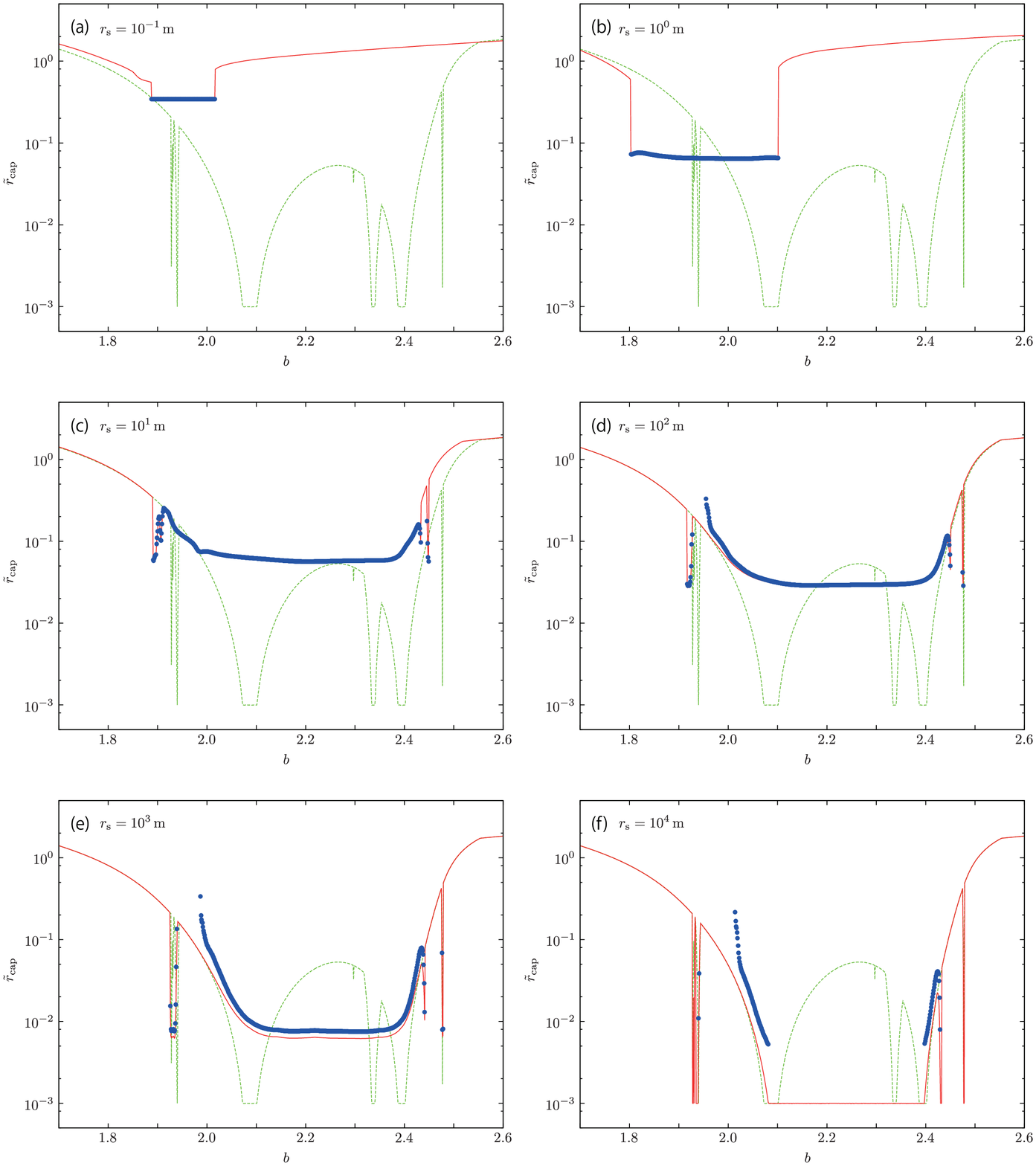}
\caption{Captured radius $\tilde{r}_{\rm cap}$ (blue) and minimum
distance from the planet until the particles are judged as captured by
the circumplanetary disk, collision with the planet, or recede enough
from the Hill sphere after encounters (red, see
\S\ref{sec:captured_radius}) as a function of $\tilde{b}$ for a wide
range of particle size ($r_{\rm s} =$ 0.1m, 1m, 10m, 100m, 1000m,
10000m).  Green curves show $\tilde{r}_{\rm min,free}$, which is the
minimum distance from the planet in the gas-free case.
\label{fig:r_cap_vs_b}}
\end{figure*}

Fig.~\ref{fig:r_cap_vs_b} shows the captured radius as a function of
$\tilde{b}$.
In the case of $r_{\rm s}=0.1$m, the particles with impact parameter
between $\tilde{b}\simeq 1.9$ and 2.0 are captured, and particles in all
the other regime basically do not enter the Hill sphere (see
Fig.~\ref{fig:orbit_small}), which is totally different behavior from
the gas-free case (green dotted line).
The position of the captured band is different from either of the two
collision bands of the gas-free case, which reflects the fact that the
motion is strongly affected by the gas flow before approaching the Hill
sphere.
In the case of $r_{\rm s}=1$m, captured band becomes wider in comparison
with the 0.1m case because the particles are easier to penetrate into
the Hill sphere through the lower-velocity higher-density region at the
post shock (see Figs.~\ref{fig:gas_flow} and \ref{fig:orbit_small}).
The captured radius is smaller than that of the 0.1m case since, to be
captured by the circumplanetary disk, larger particles need higher
density of gas and the gas density in the circumplanetary disk increases
with decreasing distance from the planet.

In the cases of $r_{\rm s}=10$m, 100m, 1000m, they show similar behavior
with some quantitative differences.
In this size regime, the particles are basically decoupled from the gas
flow at the outside of the Hill sphere, which are confirmed by the fact
that the minimum distance from the planet outside the captured band
matches well with that of the gas-free case (see red and green lines in
Fig.~\ref{fig:r_cap_vs_b}).  Width of the captured band slightly
decreases with increasing particle size, which reflects that the effect
of the gas drag for capture becomes more effective at the region closer
to the planet where gas density and relative velocity is generally
higher.
There is a flat region at the bottom of $\tilde{r}_{\rm cap}$ for each
panel.  We define the radius of the flat region as critical radius for
capture $\tilde{r}_{\rm cap,crit}$.  Once particles enter inside the
radius, gas drag is so strong that the particles are forced to move with
the gas flow of the circumplanetary disk regardless of orbits before
they reach the radius.  The typical cases for this kind of capture can
be seen in the retrograde encounter region (see
Fig.~\ref{fig:orbit_b2174}); all the particles in the retrograde
encounter region are captured by the circumplanetary disk.
But, in both edges of the captured band, we can see captured region
where $\tilde{r}_{\rm min,free}$ is larger than the critical radius.  In
this region, the particle is first captured energetically (i.e., $\tilde
{E}_{\rm J}<0$) in a highly eccentric orbit with the pericenter around
$\tilde{r}_{\rm min,free}$, and then circularized.  During the
circularization process, the particles tend to keep the pericenter, thus
$\tilde{r}_{\rm cap}$ is roughly aligned with $\tilde{r}_{\rm min,free}$
in this regime.

In the cases of $r_{\rm s}=100$m and 1000m, there is a band where
$\tilde{r}_{\rm cap} < \tilde{r}_{\rm min,free}$ around $\tilde{b} \sim
2.2$ -- 2.3.  The particles in this region approach the planet in the
retrograde direction, thus the particles cannot pass through near the
$\tilde{r}_{\rm min,free}$ as a pericenter and cannot make an elliptic
orbit like that occurred in the two edge regions.  Instead, the
particles are forced to change the direction into prograde and rotate
with disk gas that is rotating in almost Keplerian motion.

In the case of $r_{\rm s}=10000$m, there is no flat and base region for
$\tilde{r}_{\rm cap}$ because the critical radius for capture is smaller
than the planet physical radius, which means the particles collide with
the planet.  In other words, the gas drag is not strong enough to change
the direction from retrograde to prograde in the course of approaching
the planet.

In order to understand the capturing processes more deeply, we study
particle-size dependence of captured radius.
Since the capture radius is a function of $\tilde{b}$ even for single
size particles as seen in Fig.~\ref{fig:r_cap_vs_b}, we introduce
critical radius for capture $\tilde{r}_{\rm cap,crit}$ as a typical
capture radius for a given size regardless of $\tilde{b}$ so that we do
not need to consider the detail of the $\tilde{b}$ dependence.
We define $\tilde{r}_{\rm cap,crit}$ by the radius where captured radius
$\tilde{r}_{\rm cap}$ (blue dots in Fig.~\ref{fig:r_cap_vs_b}) shows
wide and flat region at the bottom of $\tilde{r}_{\rm cap}$ as seen in
Fig.~\ref{fig:r_cap_vs_b} a-e.
In order to define $\tilde{r}_{\rm cap,crit}$, we introduce two other
radii $\tilde{r}_{\rm cap,1/4}$ and $\tilde{r}_{\rm cap,1/2}$;
$\tilde{r}_{\rm cap,1/4}$ is defined so that the number of captured
particles with $\tilde{r}_{\rm cap} < \tilde{r}_{\rm cap,1/4}$ account
for 1/4 of all the captured particles under the condition of uniform
interval in $\tilde{b}$, and $\tilde{r}_{\rm cap,1/2}$ is defined in the
same way.
The width of the bottom region is roughly half of the whole captured
band, thus we define $\tilde{r}_{\rm cap,crit} = \tilde{r}_{\rm
cap,1/4}$ when $\tilde{r}_{\rm cap,1/4} \simeq \tilde{r}_{\rm cap,1/2}$.
If the difference is large, for example $\tilde{r}_{\rm
cap,1/2}/\tilde{r}_{\rm cap,1/4} > 1.1$, the $\tilde{r}_{\rm
cap,crit}$ is smaller than $\tilde{r}_{\rm p}$, and many particles are
collides with the planet, as seen in Fig.~\ref{fig:r_cap_vs_b}f and we
do not define $\tilde{r}_{\rm min,crit}$.

\begin{figure}
\epsscale{1.0}
\plotone{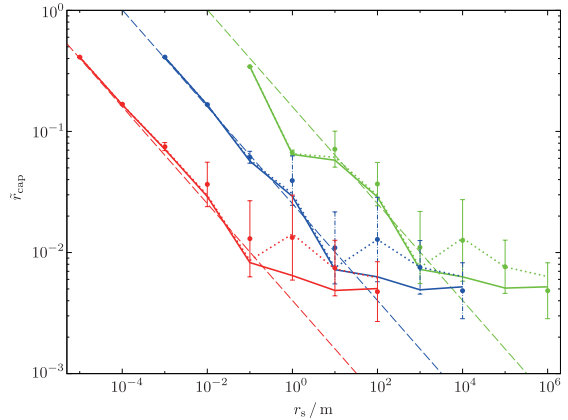}
%
\caption{Normalized capture radius as a function of particle size
 $r_{\rm s}$.  Green, blue, and red lines show scaling (depletion)
 factor of gas density $f_{\rm H} = 1$, $10^{-2}$, $10^{-4}$,
 respectively.  Solid and dotted lines show $\tilde{r}_{\rm cap,1/4}$
 and $\tilde{r}_{\rm cap,1/2}$, respectively.  Filled circles show
 $\bracket{\tilde{r}_{\rm cap}}$ (mean of $\tilde{r}_{\rm cap}$ with
 respect to $\tilde{b}$) and error bar shows standard deviation in
 logarithmic space.  Thin dashed lines show fitted lines given by
 Eq.~(\ref{r_cap_approx}).
\label{fig:r_cap_vs_r_s}}
\end{figure}

Fig. \ref{fig:r_cap_vs_r_s} shows normalized capture radii
$\tilde{r}_{\rm cap,1/4}$ and $\tilde{r}_{\rm cap,1/2}$ in the cases
with three different scaling factors of gas surface density $f_{\rm
H}=1, 10^{-2}, 10^{-4}$ (see \S~\ref{sec:C_D}).
First we can clearly see the tendency that both of the two radii
decrease with increasing particle size, which is observed in
Fig.~\ref{fig:r_cap_vs_b}.
We can also see that the difference of the two is small when the radii
are larger than $10^{-2}$.  In this regime, $\tilde{r}_{\rm cap,crit}$
can be well defined by $\tilde{r}_{\rm cap,1/4}$.
Note that the main reason why the difference between the two radii
become large at $\tilde{r}_{\rm cap} \lesssim 10^{-2}$ is the artificial
effect of the background flow; gas density at the midplane of the
circumplanetary disk in $\tilde{r}_{\rm cap} \lesssim 10^{-2}$ tends to
be smaller because of sink treatment near the origin in the hydrodynamic
simulation, in which the gas drag effect would be underestimated.

The decrease of gas density (i.e., $f_{\rm H}$) basically makes the lines
shift toward the left in the figure, because particles have to go deeper
denser region to be captured when whole gas density is uniformly
smaller.
Decrease of $f_{\rm H}$, which corresponds to gas depletion of the
protoplanetary disk due to gap formation around the planet orbit or
global disk dissipation, basically makes the lines shift toward the left
in the figure, because particles have to go deeper denser region to be
captured when whole gas density is uniformly smaller.

Note however that this is not a simple linear dependence because the gas
drag coefficient $C_{\rm D}$ is generally not a simple power-law
function of the Reynolds number and the Mach number \citep{Adachi76},
and the coefficient we use is not either (see \S \ref{sec:C_D}).  In
particular, in the case of high surface density ($f_{\rm H}=1$), gas
drag law can be Stokes regime where drag force is independent of gas
density, which can change the tendency, and which is reflected on the
jaggy curve of the case of $f_{\rm H}=1$.

We also plot fitted lines of $\tilde{r}_{\rm cap,crit}$ in the region
where $\tilde{r}_{\rm cap} \gtrsim 10^{-2}$ given by
\begin{equation}
\tilde{r}_{\rm cap}
 = 0.16
   \kakkoi{r_{\rm s}}{\rm 1m}{-0.4}
   \kakkoi{f_{\rm H}}{1}{0.4}.
\label{r_cap_approx}
\end{equation}
Although this is an empirical formula, the value 0.4 in the index can
also be estimated by the balance between energy dissipation due to gas
drag and kinetic energy of particles assuming that $C_{\rm D}$ is
constant and that particle velocity is determined only by potential
energy of the planet.  If we assume gas density is axisymmetric and the
density is described by a power-law function as $\tilde{\rho}_{\rm g}
\propto \tilde{r}^{-\gamma}$, capture radius can be analytically
obtained as $\tilde{r}_{\rm cap} \propto r_{\rm s}^{1/(1-\gamma)}$,
which is derived by comparison between dissipation energy through gas
drag and potential energy needed to be captured by the planet gravity
\citep[see][in detail]{Tanigawa10, Fujita13}.  Equating the two indexes
on $\tilde{r}_{\rm s}$, we have $\gamma = 3.5$, which is consistent with
the density distribution we use \citep{TOM12}.
We also show a simple mean radius with respect to $\tilde{b}$ given by
\begin{equation}
\bracket{\tilde{r}_{\rm cap}}
 = \exp \left(
   \frac{\displaystyle
         \int_0^\infty
             \log(\tilde{r}_{\rm cap}) \frac{3}{2}\tilde{b}d\tilde{b}}
        {\displaystyle
         \int_0^\infty
             \frac{3}{2}\tilde{b}d\tilde{b}}
        \right),
\quad
\mbox{for all the captured orbits.}
\label{eq:r_cap_ave}
\end{equation}
The mean radius $\bracket{\tilde{r}_{\rm cap}}$ also shows the similar
trend of $\tilde{r}_{\rm cap,1/2}$, but since the distribution is far
from symmetric about the mean value, the mean radius is not necessarily
suitable to define $\tilde{r}_{\rm cap,crit}$.

\subsection{Capture rate by circumplanetary disks}
\begin{figure}
\epsscale{1.0}
\plotone{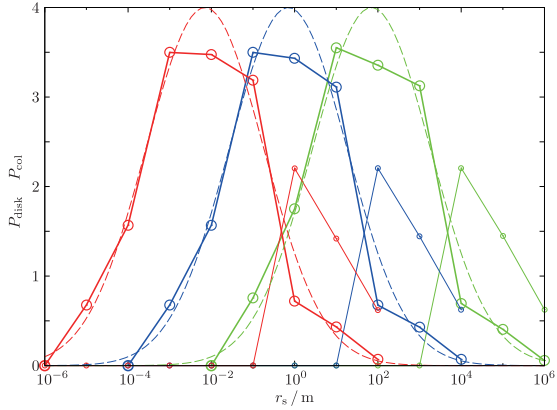}
%
\caption{Normalized probability of capture by the circumplanetary disk
 $P_{\rm disk}$ (thick lines) and the planet $P_{\rm planet}$ (thin
 lines) as a function of particle size $r_{\rm s}$.  Blue, green, red
 lines show $f_{\rm H} = 1$, $10^{-2}$, $10^{-4}$, respectively.
\label{fig:P_cap_disk_vs_r_s}}
\end{figure}

Fig. \ref{fig:P_cap_disk_vs_r_s} shows normalized probabilities of
capture by the circumplanetary disk and by the planet.

We define normalized probabilities captured by the circumplanetary disk
and the planet as
\begin{equation}
P_{\rm disk}(r_{\rm s},f_{\rm H})
 =2\int_0^\infty
       \varphi_{\rm disk}(r_{\rm s}, f_{\rm H}, \tilde{b})
       \frac{3}{2}\tilde{b} d\tilde{b},
\label{eq:P_disk}
\end{equation}
\begin{equation}
P_{\rm planet}(r_{\rm s},f_{\rm H})
 =2\int_0^\infty
       \varphi_{\rm planet}(r_{\rm s}, f_{\rm H}, \tilde{b})
       \frac{3}{2}\tilde{b} d\tilde{b},
\label{eq:P_planet}
\end{equation}
where $\varphi_{\rm disk}$ is a judgment function whether a particle is
captured by the circumplanetary disk: unity if the particle is captured,
and zero otherwise.  The definition of $\varphi_{\rm planet}$ is in the
similar way; unity if the particle collides with the planet, and zero
otherwise.  Note that all the particles that are judged as capture by
the circumplanetary disk is going to collide with the planet after long
term inward orbital evolution by gas drag, but we use the conditions for
capture described in \S \ref{sec:captured_radius}.

Although we show three different $f_{\rm H}$, we do not see significant
qualitative difference between them, so we focus on the case of $f_{\rm
H}=1$ below, unless otherwise stated.
In the limit of small particle size ($r_{\rm s} \leq 0.01$m), both
$P_{\rm disk}$ and $P_{\rm planet}$ are zero because small particles
that are well coupled with gas cannot enter into the Hill sphere.
From $r_{\rm s} = 0.1$m to 10m, $P_{\rm disk}$ increases with $r_{\rm
s}$, which corresponds to the increase of the captured band seen in
Fig.~\ref{fig:r_cap_vs_b}a-c.  But $P_{\rm planet}$ is still zero
because all the particles that enter the Hill sphere are captured by the
circumplanetary disk.
From $r_{\rm s} = 10$m to 1000m, $P_{\rm disk}$ does not change
significantly because the width of the captured band weakly decrease
with $r_{\rm s}$ as described in \S \ref{sec:captured_radius}.
When $r_{\rm s} \geq 1000$m, $P_{\rm disk}$ decreases with increasing
$r_{\rm s}$.  This is because gas drag become ineffective and some
fraction of particles collide with the planet, rather than captured by
the circumplanetary disk.
In the limit of $r_{\rm s} \rightarrow \infty$, we expect $P_{\rm disk}
= 0$ and $P_{\rm planet} = 11.3 \sqrt{\tilde{r}_{\rm p}}$ \citep{Ida89,
Inaba01}.  In the case of our setting ($\tilde{r}_{\rm p} = 10^{-3}$),
we have $P_{\rm planet} = 0.36$, in which $P_{\rm planet}$ is
approaching with increasing $r_{\rm s}$.  Note that the reason why
$P_{\rm planet} > 0.36$ in this weak-drag regime is that gas drag
enhances the collision rate onto the planet \citep{Inaba03,Tanigawa10}.

Finally we fit $P_{\rm disk}$ by an formula.  An empirical formula for
$P_{\rm disk}$ can be roughly approximated as
\begin{align}
&P_{\rm disk}(r_{\rm s}, f_{\rm H}) \nonumber \\
&=
\begin{cases}
 P_{\rm max}
 \exp \left[ -\left(\displaystyle
                    \frac{\log (r_{\rm s}/r_{\rm s,peak})}
                         {\log W_{\rm s,HWHM}}
              \right)^2
      \right]
 & \mbox{if $r_{\rm s} \gtrsim 5\e{-4} r_{\rm s,peak}$}, \\
 0
 & \mbox{otherwise},
\end{cases}
\label{Pcap_disk_approx}
\end{align}
where $P_{\rm max} = 4.0$, $r_{\rm s,peak} = 70f_{\rm H}$ m, $W_{\rm
s,HWHM} = 100$.  The fitted lines are also plotted in
Fig.~\ref{fig:P_cap_disk_vs_r_s}.  This formula is not derived by
physical consideration, but it might be useful for rough estimation.

\begin{figure}
\epsscale{1.0}
\plotone{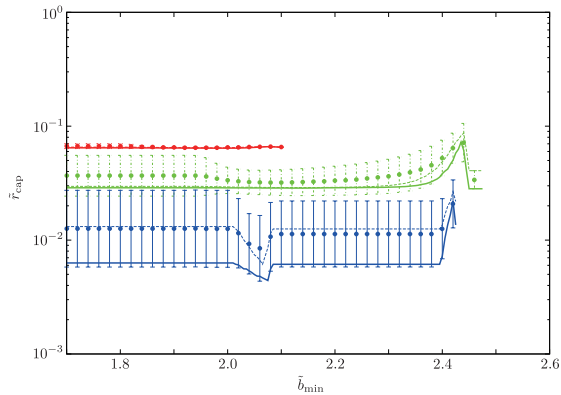}
%
\caption{Dependence of mean capture radius $\tilde{r}_{\rm cap}$ on
width of particle gap.  Horizontal axis is $\tilde{b}_{\rm min}$, which
is the lower bound of the region where particles exist.  Red, green and
blue show the case with $r_{\rm s} = 1, 10^2, 10^4$ m particles,
respectively.  Filled circles show log average and error bars show one
sigma.  Thick lines and thin dashed lines show $\tilde{r}_{\rm cap,1/4}$
and $\tilde{r}_{\rm cap,1/2}$, respectively.
\label{fig:r_cap_vs_b_min}}
\end{figure}

\begin{figure}
\epsscale{1.0}
\plotone{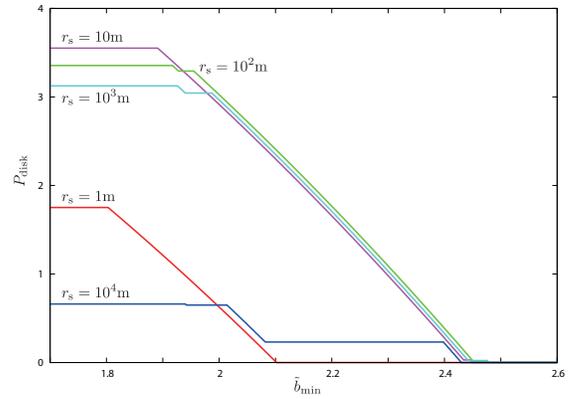}
\caption{Dependence of normalized probability of the capture by the
 circumplanetary disk on the width of particle gap.
\label{fig:P_cap_disk_vs_b_min}}
\end{figure}


\section{Discussion}
%
We have assumed so far that particle surface density is uniform in the
protoplanetary disk before particles approach the planet, but that is
not true in general.
In particular, a particle gap, which is a lower surface density annular
region round the planet orbit, can form easier than the gap of gas
\citep[e.g.,][]{Tanaka97, Paardekooper07, Zhou07, Shiraishi08, Ayliffe12}.
We will examine the effect of the gap opening on the accretion rate of
the particles.
To examine the effect, we calculate $\tilde{r}_{\rm
cap}$ and $P_{\rm disk}$ as a function of gap width that we define by
$\tilde{b}_{\rm min}$ so that particles uniformly exist at $\tilde{b} >
\tilde{b}_{\rm min}$ and there are no particles at $\tilde{b} <
\tilde{b}_{\rm min}$.
Fig.~\ref{fig:r_cap_vs_b_min} shows the dependence of $\tilde{r}_{\rm
cap,1/2}$, $\tilde{r}_{\rm cap,1/4}$, and $\bracket{\tilde{r}_{\rm
cap}}$ on $\tilde{b}_{\rm min}$.
In the case with $r_{\rm s} = 1$m, $\tilde{r}_{\rm cap}$ do not depend
on $\tilde{b}_{\rm min}$ almost at all, and even when $r_{\rm s} = 10^2$
and $10^4$ m cases, $\tilde{r}_{\rm cap}$ changes only by a factor of a
few.
This shows that even when the particle gap is formed and particles
distribution is far from uniform, there is no significant impact on
capture radius.

Fig.~\ref{fig:P_cap_disk_vs_b_min} shows $P_{\rm disk}$ as a function of
$\tilde{b}_{\rm min}$ for various values of $r_{\rm s}$.  In contrast to
Fig.~\ref{fig:r_cap_vs_b_min}, we can see that $P_{\rm disk}$ decreases
almost linearly with $\tilde{b}_{\rm min}$ in the region where particles
are captured.
This can be easily understood by Fig.~\ref{fig:r_cap_vs_b}.
This simply means that when particle gap opens widely, accretion rate
onto the circumplanetary disk reduces, and when $\tilde{b}_{\rm min}
\gtrsim 2.4$, no particle accretion is expected.
Note that the holizontal parts of the lines indicate that non-capture
regions such as $1.93 < \tilde{b} < 1.98$ in Fig.~\ref{fig:r_cap_vs_b}e or
$2.09 < \tilde{b} < 2.40$ in Fig.~\ref{fig:r_cap_vs_b}f.

\citet{Muto09} derived an analytic formula that describes radial
migration of small particles near a low-mass planet embedded in a
protoplanetary disk.
According to Eq.~(68) of their paper and comparing the two dominant
terms (gravitational scattering by the planet and radial inward
migration due to slight difference of rotation velocities), we obtain
the gap width, which corresponds to $\tilde{b}_{\rm min}$ in this study,
as 2.04, in the case that normalized stopping time is unity and degree
of non-Keplerian rotation of disk gas ($\eta$ in their notation) is
$10^{-3}$.
This would mean that particle gap is still narrow enough for particles
to accrete onto circumplanetary disks (see
Fig.~\ref{fig:P_cap_disk_vs_b_min}).
Note however that we extrapolate their formula beyond their assumption
(i.e., they do not consider large density change, such as gap formation
of gas disk), which would probably change the estimation here.

Although gap structure of gas would create a particle gap and dam radial
flow of particles toward the planet, particles in a particular size
range can pass through the gap and be able to approach the planet
\citep{Rice06, Paardekooper07, Ward09, Morbidelli12, Zhu12}.
Also, strong pressure gradient at the gap edge of gas generates
hydrodynamic instability such as Rayleigh instability
\citep{Chandrasekhar61, Papaloizou84}, Rossby wave instability
\citep{Li01, Lin13}, and baloclinic instability \citep{Klahr03}, which
generate vortex and disturb the gas flow, which promote particle
diffusion in radial direction, then particles can approach to the
planet.
The dynamics at the gap edge with particles has not been well
understood, so detailed investigation on the gap dynamics is needed to
understand solid accretion onto circumplanetary disks and resultant
satellite formation.

Recently, \citet{Fujita13} has investigated motion of planetesimals in
heliocentric orbits in order to examine whether the planetesimals are
captured by the circumplanetary disk of giant planets.
They focus on planetesimals with size larger than that of ours, which
means that gas drag is weak.
They assume that the circumplanetary disk is axisymmetric around the
planet and hydrostatic equilibrium in the direction perpendicular to the
disk central plane, which is justified by their setting of large size
objects.
Although they do not obtain capture radius which we show in this paper,
they consider non-zero initial eccentricity and inclination for the
approaching objects.
\citet{Fujita13} and our work are thus in a mutually complementary
relationship, and future works along the line of these studies will
provide better understandings of satellite formation processes.

In this study, we observe that captured particles are rotating in
prograde direction, and \citet{Johansen10} also showed that particles of
a few cm in radius are rotating in prograde direction around
protoplanets of a few hundred kilometers when the particles are captured
by the protoplanets.
Since particle density is much higher than gas density, particle motion
seems to determine the rotating direction, whereas particles are dragged
by gas that is rotating in prograde direction in our case.
Although there is huge difference in mass for the two cases, we can
observe common physical property that objects in a rotating frame tend
to rotate in the same direction as the frame rotation by Coriolis force
when they are pulled toward the center, as in tropical cyclones.

We have examined how particles in heliocentric orbits are captured by
circumplanetary disks,
but the captured particles, which are rotating in the similar velocity
to that of the circumplanetary disk gas, are still migrating inward
because of slight difference of the rotation velocities between gas and
particles.
This inward drift of particles in circumplanetary disks is important in
the context of satellite formation because when the accretion rate of
particles into the circumplanetary disks, which we have obtained, is
given, the radial velocity determines the surface density of solid,
which would then determine satellite growth rate.
Assuming axisimmetric and isothermal for the circumplanetary disk, we
can obtain rotation velocity of gas, gas drag force acting on the
particles, and then inward migration velocity for the particles
\citep{Weidenschilling77,Nakagawa86}.
For example, inward velocity for 1m particles is about 5m\,s$^{-1}$ at
0.01 Hill radius from the planet, which corresponds to at $\sim 7R_{\rm
J}$ for a planet at 5AU.
Applying the accretion rate given by Eq.~(\ref{Pcap_disk_approx}) and
assuming steady state inward particle flow, we can estimate solid
surface density as 1g\,cm$^{-2}$, which might be a bit small for
satellite formation.
However, the solid surface density estimated depends on particle size
and gas density (which corresnponds to $f_{\rm H}$ in this paper), and
the drag law itself depends on the two parameters.  Thus these
dependences have to be examined in the future.
In addition, size of particles in heliocentric orbits near giant planets
is important for satellite formation processes because it affects
accretion rate obtained in this study and the filtering effect for
particles at the edge of gas gap produced by the giant planet.
A recent statistical method that uses a coagulation equation with
fragmentation showed that a large amount of particles of 1-100m in size
are produced by fragmentation \citep{Kobayashi12}.
A comprehensive circumplanetary disk model that considers size
distribution of incoming particles and growth in the disk would be
necessary in the future in order to understand more realistic satellite
formation processes.

%
\citet{Crida12} has recently proposed a totally different mechanism to
reproduce the regular satellites.
They considered a heavy and compact ring composed of small particles.
Diffusion processes in the ring make it spread outward, and once
particles are transported beyond the Roche limit, they are allowed to
accumulate gravitationally to be a larger clump, which is a
proto-satellite.
The proto-satellite moves outward through tidal interaction with the
planet and the ring, and once the proto-satellite migrates far enough,
the second proto-satellite start to form.
The ring produces many proto-satellites in this way.
However, tidal interactions of outer (older) satellites are weaker, they
migrate slower than inner ones, and tend to be captured by inner ones,
which leads that outer satellites tend to be larger.
Since this ``pyramidal'' size distribution is consistent with the
current icy satellites around Saturn, Uranus, and Neptune, this
mechanism would be likely to have occurred.
This scenario need to have a heavy ring around the planet.
\citet{Estrada06} proposed a mechanism to supply solid materials into
the Hill radius by collision between heliocentric planetesimals under
gas-free condition, which may help to have a ring around the planet.
In addition, this mechanism cannot explain Galilean satellites.
Both mechanisms, formation from a gas disk with solid and formation from
a ring without gas, have their advantages and disadvantages, so we may
have to consider hybrid scenarios to explain the formation process of
the current satellite systems.

\section{Conclusions}
We have demonstrated how solid particles in heliocentric orbits are
captured by a circumplanetary disk around an actively growing giant
planet embedded in a protoplanetary disk by using numerical integration
of particle orbits with gas drag.
We found that
distance from the planet (orbital radius around the planet) when the
particle is captured by the circumplanetary disk decreases with
increasing particle size.  The captured radius is approximated by a
fitting function Eq.~(\ref{r_cap_approx}).
The main contribution to the accretion is the regime where particles
encounter with the planet in regrograde direction, which corresponds to
the regrograde encounter regime in $\tilde{b}$ space
(Fig.~\ref{fig:rmin_vs_b}).
We also found that
the accretion efficiency is maximum when the size is $\sim 10^2$m in the
case of the surface density of the minimum mass solar nebula and 5AU
planet.
Width of the profile of normalized capture probability with respect to
size is wide even in log scale (about two-order of magnitude in size).
If the size is smaller than a critical size, particles cannot accrete
onto the circumplanetary disk because of strong coupling with gas, which
cannot accrete through the midplane even when active gas accretion
phase.
The size dependence of the accretion efficiency is approximated by
Eq.~(\ref{Pcap_disk_approx}).
Even when a particle gap around the planet orbit is formed, captured
radius is hardly affected by the gap, but accretion rate would be
reduced and could be zero depending on the gap width.
Several studies on the formation of particle gaps have been done.
In particular, particle motion is strongly affected by the motion of
gas, and the structure of the {\it gas} gap was not well understood at
this stage mainly because the gas gap structure is affected by some
hydrodynamic instability.
Effect of the particle gap is important for satellite formation, and
thus more studies on gas and particle gap should be done in the future.

%
%
%


\acknowledgments
We are grateful to Hidekazu Tanaka, Keiji Ohtsuki, Hiroshi Kobayashi,
and Satoshi Okuzumi, Taku Takeuchi, Alessandro Morbidelli, Aurelian
Crida for their valuable comments.
We also thank the referee for comments that improve the manuscript.
T.T. is supported by Grant-in-Aid for Scientific Research (23740326 and
24103503) from the MEXT of Japan.
M.N.M. is supported by Grant-in-Aid for Scientific Research (25400232)
from the MEXT of Japan.
This work was supported by Center for Planetary Science running under
the auspices of the MEXT Global COE Program entitled ``Foundation of
International Center for Planetary Science''.
Numerical calculations were carried out on NEC SX-9 at Center for
Computational Astrophysics, CfCA, of National Astronomical Observatory
of Japan.
A part of the figures were produced by GFD-DENNOU Library.

\clearpage


\begin{thebibliography}{}
\bibitem[Adachi et al.(1976)]{Adachi76} Adachi, I., Hayashi, C., \&
		Nakazawa, K. 1976, Prog. Theor. Phys., 56, 1756
\bibitem[Ayliffe \& Bate(2009)]{Ayliffe09b} Ayliffe, B. A., \& Bate,
		M. R. 2009, \mnras, 397, 657
\bibitem[Ayliffe et al.(2012)]{Ayliffe12} Ayliffe, B. A., Laibe, G.,
		Price, D. J., \& Bate, M. R. 2012, \mnras, 423, 1450
\bibitem[Bate et al.(2003)]{Bate03}  Bate, M. R., Lubow, S. H., Ogilvie,
 G. I., \& Miller, K. A. 2003, \mnras, 341, 213
\bibitem[Bodenheimer \& Pollack(1986)]{Bodenheimer86} Bodenheimer, P.,
 \& Pollack, J. B. 1986, \icarus, 67, 391
\bibitem[Canup \& Ward(2002)]{Canup02} Canup, R. M., \& Ward, W. R. 2002,
 \aj, 124, 3404
\bibitem[Canup \& Ward(2006)]{Canup06} Canup, R. M., \& Ward, W. R. 2006,
 \nat, 441, 834
\bibitem[Champman \& Cowling(1970)]{CC70} Champman, S., \& Cowling,
		T. G. 1970, Cambridge: University Press, 1970, 3rd ed.
\bibitem[Chandrasekhar(1961)]{Chandrasekhar61} Chandrasekhar, S. 1961,
 Hydrodynamic and Hydromagnetic Stability, Clarendon press
\bibitem[Crida \& Charnoz(2012)]{Crida12} Crida, A., \& Charnoz,
		S. 2012, Science, 228, 1196
\bibitem[D'Angelo et al.(2002)]{DAngelo02} D'Angelo, G., Henning, T., \&
 Kley, W. 2002, \aap, 385, 647
\bibitem[D'Angelo et al.(2003)]{DAngelo03} D'Angelo, G., Kley, W., \&
 Henning, T. 2003, \apj, 586, 540
\bibitem[Estrada \& Mosqueira(2006)]{Estrada06} Estrada, P., \&
		Mosqueira, I. 2006, \icarus, 181, 486
\bibitem[Estrada et al.(2009)]{Estrada09} Estrada, P.,
		Mosqueira, I., Lissauer, J. J., D'Angelo, G., \&
		Cruikshank, D. P. 2009, in Europa, ed. R. T. Pappalardo,
		W. B. McKinnon, \& K. Khurana (Tucson, AZ: Univ. Arizona
		Press), 27
\bibitem[Fujita et al.(2013)]{Fujita13} Fujita, T., Ohtsuki, K.,
		Tanigawa, T., and Suetsugu, R. 2013, \aj, 146, 140
\bibitem[Giuli(1968)]{Giuli68} Giuli, R. T. 1968, \icarus, 8, 301
\bibitem[Gressel et al.(2013)]{Gressel13} Gressel, O., Nelson, R. P.,
		Turner, N. J., \& Ziegler, U. 2013, \apj, 779, 59
\bibitem[Hayashi(1985)]{Hayashi85} Hayashi, C., Nakazawa, K., \&
		Nakagawa, Y. 1985, Protostars and planets II, 1100
\bibitem[Henon \& Petit(1986)]{Henon86} Henon, M., \& Petit,
		J.-M. 1986, Celestial Mechanics, 38, 67
\bibitem[Ida \& Nakazawa(1989)]{Ida89} Ida, S., \& Nakazawa, K. 1989,
		\aap, 224, 303
\bibitem[Ida(1990)]{Ida90} Ida, S. 1990, \icarus, 88, 129
\bibitem[Ikoma et al.(2000)]{Ikoma00} Ikoma, M., Nakazawa, K., \& Emori,
		H. 2000, \apj, 537, 1013
\bibitem[Inaba et al.(2001)]{Inaba01} Inaba, S., Tanaka, H., Nakazawa,
		K., Wetherill, G. W., \& Kokubo, E. 2001, \icarus, 149,
		235
\bibitem[Inaba \& Ikoma(2003)]{Inaba03} Inaba, S., \& Ikoma, M. 2003,
		\aap, 410, 711
\bibitem[Iwasaki \& Ohtsuki(2007)]{Iwasaki07} Iwasaki, K. \& Ohtsuki,
		K. 2007, \mnras, 377, 1763
\bibitem[Johansen \& Lacerda(2010)]{Johansen10} Johansen, A. \& Lacerda,
		P. 2010, \mnras, 404, 475
\bibitem[Kley(1999)]{Kley99} Kley, W. 1999, \mnras, 303, 696
\bibitem[Kary \& Dones (1996)]{Kary96} Kary, D. M., \& Dones, L. 1996,
		\icarus, 121, 207
\bibitem[Klahr \& Bodenheimer(2003)]{Klahr03} Klahr, H. \& Bodenheimer,
		P. 2003, \apj, 582, 869
\bibitem[Klahr \& Kley(2006)]{Klahr06} Klahr, H. \& Kley, W. 2006, \aap,
		445, 747
\bibitem[Kobayashi et al.(2012)]{Kobayashi12} Kobayashi, H., Ormel,
		C. W., \& Ida, S. 2012, \apj, 756, 70
		H., \& Krivov, A. V. 2011, \apj, 738, 35
\bibitem[Korycansky \& Papaloizou (1996)]{Korycansky96} Korycansky,
		D. G., \& Papaloizou, J. C. B. 1996, \apjs, 105, 181
\bibitem[Li et al.(2001)]{Li01} Li, H., Colgate, S. A., Wendroff, B., \&
		Liska, R. 2001, 551, 874
\bibitem[Lin(2013)]{Lin13} Lin, M. 2013, \apj, 765, 84
\bibitem[Lubow et al.(1999)]{Lubow99} Lubow, S. H., Seibert, M., \&
    Artymowicz, P. 1999, \apj, 526, 1001
\bibitem[Lunine \& Stevenson(1982)]{Lunine82} Lunine, J. I., \&
		Stevenson, D. J. 1982, \icarus, 52, 14
\bibitem[Machida et al.(2005)]{MachidaMTH05} Machida, M. N., Matsumoto,
		T., Tomisaka, K., \& Hanawa, T. 2005, \mnras, 362, 369
\bibitem[Machida et al.(2008)]{Machida08} Machida, M. N., Kokubo, E.,
		Inutsuka, S., \& Matsumoto, T. 2008, \apj, 685, 1220
\bibitem[Machida et al.(2010)]{Machida10} Machida, M. N., Kokubo, E.,
		Inutsuka, S., \& Matsumoto, T. 2010, \mnras, 405, 1227
\bibitem[Matsumoto \& Hanawa(2003)]{Matsumoto03b} Matsumoto, T., \&
		Hanawa, T. 2003, \apj, 595, 913
\bibitem[Miki(1982)]{Miki82} Miki, S. 1982, Prog. Theor. Phys., 67, 1053
\bibitem[Mizuno(1980)]{Mizuno80} Mizuno, H. 1980, Prog. Theor. Phys., 64, 544
\bibitem[Morbidelli \& Nesvorny(2012)]{Morbidelli12} Morbidelli, A., \&
		Nesvorny, D. 2012, \aap, 546, 18
\bibitem[Mosqueira \& Estrada(2003)]{Mosqueira03a} Mosqueira, I., \&
		Estrada, P. R. 2003, \icarus, 163, 198
\bibitem[Muto \& Inutsuka(2009)]{Muto09} Muto, T., \& Inutsuka, S. 2009,
		\apj, 695, 1132
\bibitem[Nakagawa et al.(1986)]{Nakagawa86} Nakagawa, Y.,
		Sekiya, M., \& Hayashi, C. 1986, \icarus, 67, 375
\bibitem[Nakazawa \& Ida(1988)]{Nakazawa88} Nakazawa, K., \& Ida,
		S. 1988, Prog. Theor. Phys. Suppl., 96, 167
\bibitem[Nishida(1983)]{Nishida83} Nishida, S. 1983,
		Prog. Theor. Phys. 70, 93
\bibitem[Ohtsuki(1999)]{Ohtsuki99} Ohtsuki, K. 1999, \icarus, 137, 152
\bibitem[Ohtsuki et al.(2002)]{Ohtsuki02} Ohtsuki, K., Stewart, G. R.,
		\& Ida, S. 2002, \icarus, 155, 436
\bibitem[Okuzumi et al.(2012)]{Okuzumi12} Okuzumi, S., Tanaka, H.,
		Kobayashi, H., \& Wada, K. 2012, \apj, 752, 108
\bibitem[Paardekooper(2007)]{Paardekooper07} Paardekooper, S.-J. 2007,
		\aap, 462, 355
\bibitem[Paardekooper \& Mellema(2008)]{Paardekooper08} Paardekooper,
		S.-J., \& Mellema, G. 2008, \aap, 478, 245
\bibitem[Papaloizou and Pringle(1984)]{Papaloizou84} Papaloizou,
		J. C. B., \& Pringle J. E. 1984, \mnras, 208, 721
\bibitem[Petit \& Henon(1986)]{Petit86} Petit, J. M., \& Henon, M.,
		1986, \icarus, 66, 536
\bibitem[Press et al.(2007)]{numerical_recipes} Press, W. H., Teukolsky,
		S. A., Vetterling, W. T., \& Flannery, B. P. 2007,
		Numerical Recipes 3rd Edition: The Art of Scientific
		Computing
\bibitem[Rice et al.(2006)]{Rice06} Rice, W. K. M., Armitage, P. J.,
		Wood, K., and Lodato, G. 2006, \mnras, 373, 1619
\bibitem[Sekiya et al.(1987)]{Sekiya87} Sekiya, M., Miyama,
		S., \& Hayashi. C. 1987, EM\&P, 39, 1
\bibitem[Shakura \& Sunyaev(1973)]{Shakura73} Shakura, N. I., \&
		Sunyaev, R. A. 1973, \aap, 24, 337
\bibitem[Shiraishi \& Ida(2008)]{Shiraishi08} Shiraishi, M., \& Ida,
		S. 2008, \apj, 684, 1416
\bibitem[Szulagyi et al.(2014)]{Szulagyi14} Szulagyi, J., Morbidelli,
		A., Crida, A., \& Masset, E. 2014, accepted for
		publication in \apj
\bibitem[Tanaka \& Ida(1997)]{Tanaka97} Tanaka, H., \& Ida, S. 1997,
		\icarus, 125, 302
\bibitem[Tanigawa \& Watanabe(2002)]{Tanigawa02} Tanigawa, T., \&
		Watanabe, S. 2002, \apj, 580, 506
\bibitem[Tanigawa \& Ohtsuki(2010)]{Tanigawa10} Tanigawa, T., \& Ohtsuki,
		K. 2010, \icarus, 205, 658
\bibitem[Tanigawa et al.(2012)]{TOM12} Tanigawa, T., Ohtsuki, K., \&
		Machida, M. N. 2012, \apj, 747, 47
\bibitem[Watanabe \& Ida(1997)]{Watanabe97} Watanabe S., \& Ida,
		S. 1997, Sec. 3 of Comparative Study of Planetology,
		Vol. 12 of Earth and Planetary Science series (in
		Japanese), Iwanami Shoten Publishers
\bibitem[Ward(2009)]{Ward09} Ward, W. R. 2009, in Lunar and Planetary
		Institute Science Conference Abstracts, the Woodlands,
		Texas: Lunar and Planetary Science, 40, 1477
\bibitem[Ward \& Canup(2010)]{Ward10} Ward, W. R., \& Canup, R. M. 2010,
		\aj, 140, 1168
\bibitem[Weidenschilling(1977)]{Weidenschilling77} Weidenschilling,
		S. J. 1977, \mnras, 180, 57
\bibitem[Youdin \& Lithwick(2007)]{Youdin07} Youdin, A. N., \& Lithwick,
		Y. 2007, \icarus, 192, 588
\bibitem[Zhou and Lin(2007)]{Zhou07} Zhou, J., \& Lin, D. N. C. 2007,
		\apj, 666, 447
\bibitem[Zhu et al.(2012)]{Zhu12} Zhu, Z., Nelson, R. P., Dong, R.,
		Espaillat, C., \& Hartman, L. 2012, \apj, 755, 6
\end{thebibliography}
\end{document}